\documentclass[aps,pre,reprint,doublespace]{revtex4-1}%
\usepackage{amsmath}%
\usepackage{amsfonts}%
\usepackage{amssymb}%
\usepackage{graphicx}
\usepackage{epstopdf}
\usepackage{natbib}
\begin{document}
\title{Network Overload due to Massive Attacks}
\author{Yosef Kornbluth$^{1,2}$, Gilad Barach$^{1,3}$, Mark Tuchman$^{1,4}$, Benjamin Kadish$^1$, Gabriel Cwilich$^1$, and Sergey V. Buldyrev$^1$\\
$^1$ Department of Physics, Yeshiva University, 500 West 185th Street, New
York, New York 10033\\
$^2$ Massachusetts Institute of Technology, Department of Mechanical Engineering, Cambridge, MA 02135\\
$^3$ Sixdof Space, Yad Harutsim 19, Jerusalem, Israel \\
$^4$ Stanford University, Department of Material Science, Stanford University, 450 Sierra Mall, Stanford, CA 94305}

\begin{abstract}
  We study the cascading failure of networks due to overload, using the  betweenness centrality of a node as the measure of its load following the Motter and Lai model. We study the fraction of survived nodes at the end of the cascade $p_f$ as function of the strength of the initial attack, measured by the fraction of nodes $p$, which survive the initial attack for different values of tolerance $\alpha$ in random regular and Erd\"os-Renyi graphs. We find the existence of first order phase transition line $p_t(\alpha)$ on a $p-\alpha$ plane, such that if $p <p_t$ the cascade of failures lead to a very small fraction of survived nodes $p_f$ and the giant component of the network disappears, while for $p>p_t$, $p_f$ is large and the giant component of the network is still present. Exactly at $p_t$ the function $p_f(p)$ undergoes a first order discontinuity. We find that the line $p_t(\alpha)$ ends at critical point $(p_c,\alpha_c)$ ,in which the cascading failures are replaced by a second order percolation transition. We analytically find the average betweenness of nodes with different degrees before and after the initial attack, investigate their roles in the cascading failures, and find a lower bound for $p_t(\alpha)$. We also study the difference between a localized and random attacks.
\end{abstract}
\date{\today}
\maketitle

\section{Introduction}
In August 2003, a power failure struck northeastern North America and 55 million people lost power. It is commonly accepted that the cause of this event was a series of cascading failures in the power grid \cite{Albert}. A failure in one part of the network causes that some region of the system to be overloaded and this then causes other parts of the network to fail. This process can repeat multiple times until a large portion of the network has failed. In the case of the Northeastern power grid, this process resulted in a widespread blackout.\\
To explore this phenomenon, we use a model developed by Motter and Lai \cite{motterLai,motter}. They study the betweenness of a node, defined as the number of the shortest paths connecting any pair of nodes in the network that pass through (but do not end in) this node. A network is constructed, and we calculate the initial betweenness $b^{(o)}_i$ of each node $i$. A node can withstand a maximum betweenness of $L_i\equiv (1+\alpha)b^{(o)}_i$, where  $\alpha$, the tolerance, is a global parameter of the system. A fraction $(1-p)$ of nodes is removed, and the betweenness of the surviving nodes  is recalculated. The nodes whose betweenness $b_i$ is greater than $L_i$ are destroyed and removed from the network. The betweenness of the surviving nodes is again recalculated, and the nodes whose new betweenness exceed  $L_i$ are removed. This process is repeated until no more nodes fail due to overload and we find the fraction of survived nodes $p_f(p,\alpha)<p$.  We find that function $p_f(p,\alpha)$ has a first order discontinuity at $p=p_t(\alpha)$. Above this point, the network is intact, and a majority of the surviving nodes are part of a giant component $P_\infty$. The rest of the survived nodes are isolated from the giant component; because they connect to fewer nodes, they will have a very low betweenness and, furthermore, will not contribute to the betweenness of the nodes of the giant component. Although these nodes technically survive, they do not contribute to the global connectivity of the network. Thus, we will often focus  only on the size of the giant component $P_\infty$, rather than the total number of surviving nodes. If $p<p_t$, the giant component disappears, but the fraction of survived nodes $p_f$ is still finite.\\ 
Most of the research until now has explored the effects of the failure of a single node~\cite{Wang,Crucitti,Zhao,Zhao2, Wang2}. We will study numerically the effects of a massive attack on the network, exploring the values of the parameters which lead to the network's collapse and the nature of that collapse, also using analytical insights.
 In the real world, this massive attack could come from a natural disaster or a human attack on a nation's infrastructure.\\
We study the behavior of the network when the size of the attack is close to the "threshold attack" $p_t$.   For initial attacks $(1-p)<(1-p_t )$  the network will survive with a majority of its nodes intact, while for $(1-p)>(1-p_t )$  it will disintegrate.  The network will fail approximately when the nodes that end up with the highest betweenness after the initial attack have a betweenness that is near that of their limit. At that point, the failure of a single node will redistribute the "load" of that node such that one or more other nodes will fail in turn.   This attack, then, creates the conditions for a cascade, triggering a sequence or cascade of failures that will not end until the network is destroyed. 
\section{Numerical Results of the Threshold Point}

\begin{figure}
\includegraphics[clip,width=\columnwidth]{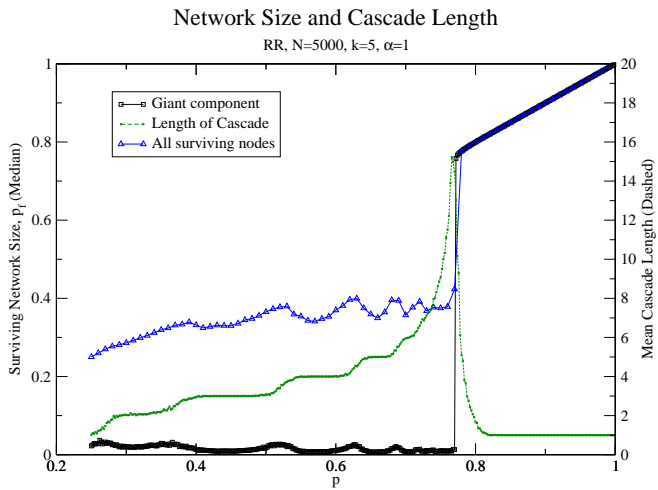}
\caption{Median surviving network size and number of cascades for a random regular graph as a function of $p$. The data presented is the result of averaging over 500 realizations.  Note that the number of cascades peaks just at the first-order transition, where the size of the surviving network drops suddenly, at $p_t=.771$ for $N=5000$, $k=5$, and $\alpha=1$.  We present data for both the size of the largest surviving component of the system and the number of nodes that do not fail due to overload or the initial attack, including isolated nodes. The significance of these two quantities is discussed in the introduction.} \label{f:cascpeak}
\end{figure}

For sufficiently low tolerances we find that, as a function of the size of the initial attack $(1-p)$,  the behavior of the network experiences a first order phase transition at a value of $p$ denoted as  $p_t$, in which the destruction of even a single additional node can trigger a cascade of failures that causes a network to collapse  (Fig. \ref{f:cascpeak}).  The principal characteristics of the first-order phase transition is the bimodality of the distribution of the order parameter, which can be either the fraction of surviving nodes or the fraction of nodes in the giant component at the end of the cascade of failures. For these first-order transitions, we can numerically find $p_t$ as the value at which the areas of both peaks, corresponding to large and small fractions of surviving nodes, are equal to each other\cite{Lowinger}.  This coincides with the value of $p$ at which  the average length of the cascade reaches a maximum.\cite{Buldyrev}\\ 
The steps in the  cascade length, and the associated fluctuations in the number of survived nodes for$p<p_t$ are caused by the discreteness of the number of cascades neccessary to approach the percolation transition of the network starting from the initial fraction of survived nodes, $p$. If after the $n$-th step $p_n$ is still larger than the percolation transition, a giant component may still exist, but its size is small enough for the betweennees of its members to be still below the maximal betweeenness. As $p$ increases, $p_n$ also increases, the size of the giant component increases and some of its memberes may exceed the maximal betweenness. At this point an additional, $n+1$ step may become necessary, and the average $p_f$ will be in between the large $p_n$ and the small $p_{n+1}$ and starts to decrease together with the giant component, while average number of cascades starts to increase from $n$ to $n+1$.

\begin{figure}
\includegraphics[clip,width=\columnwidth]{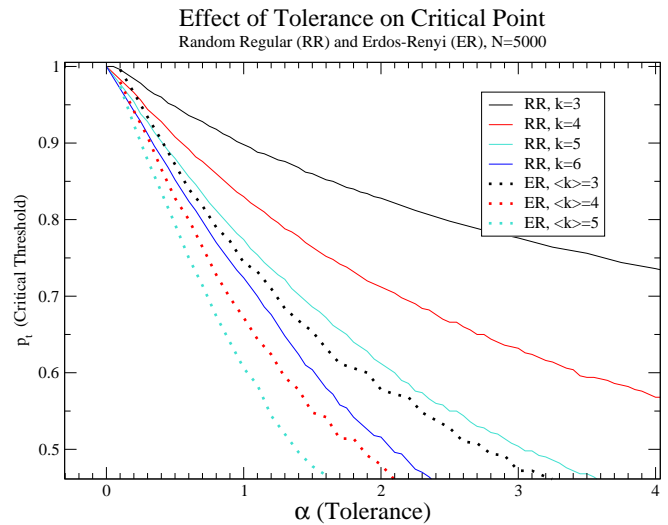}
\caption{We show $p_t$, the minimum initial survivability of a catastrophic attack, as a function of  the tolerance $\alpha$  in both random regular graphs and Erd\"os-Renyi graph. Note that for identical conditions, the regular graph is more resilient; as  is discussed in the text, the variation in initial degree in the nodes of the Erd\"os-Renyi graphs causes some nodes to be more susceptible to failure. The data plotted were obtained by studying the results of 50 realizations for each set of parameters.}
\label{f:pcVsaER}
\end{figure}

We studied how the value of the size of the threshold initial attack $p_t$ depends on the different values of the tolerance $\alpha$ for graphs with different connectivity.  In the case of random regular graphs (RR), we show data for different values of the degree $k$.  In the case of  Erd\"os-Renyi graphs  (ER) we present data for different values of the average degree $\langle k\rangle$ (Fig. \ref{f:pcVsaER}). It can be seen, as we would expect, that as the tolerance increases, the network becomes more resilient and $p_t$ decreases. This feature is common to both types of networks. For the same tolerance, the ER graphs with the average degree $\langle k\rangle$ are in all cases more resilient than the RR graphs with degree $k=\langle k\rangle$. 
We will show later that at sufficiently high tolerances, the collapse of the network changes its nature, and we observe a more gradual second-order transition.
\section{General Results}
\label{gen_res}
\begin{figure}
\includegraphics[clip,width=\columnwidth]{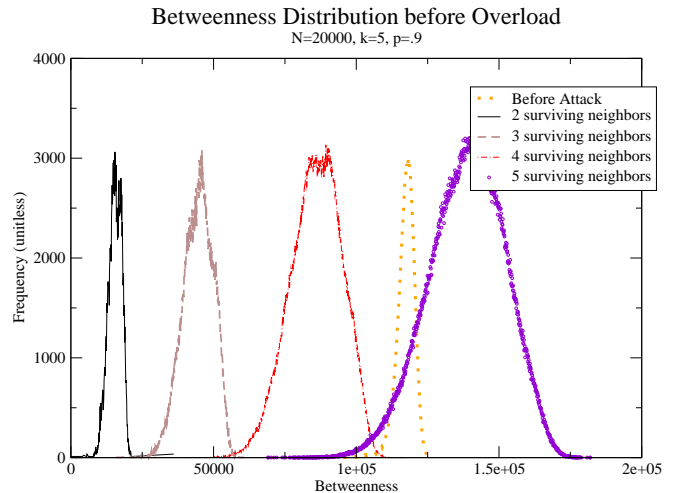}
\caption{Betweenness distribution after the initial attack, but before any failure due to overload, for RR graphs with $N=20000$ nodes, $k=5$, and $p=0.9$. Each curve represents the distribution of betweenness for a different number of surviving neighbors, $\ell$, normalized so that the peaks are all the same height. The betweenness distribution before the initial attack is included for comparison.  It is sharply peaked around its analytical prediction $\langle B\rangle =N\ln (N/k)/\ln(k-1) \approx 1.20\times 10^5$. These results are the combination of 100 realizations. The position of the peaks of the curves are approximately in a 2:6:12:20 ratio, showing a dependence on $\ell(\ell-1)$
}
\label{f:betAfter}
\end{figure}

\begin{figure}
\includegraphics[clip, width=\columnwidth]{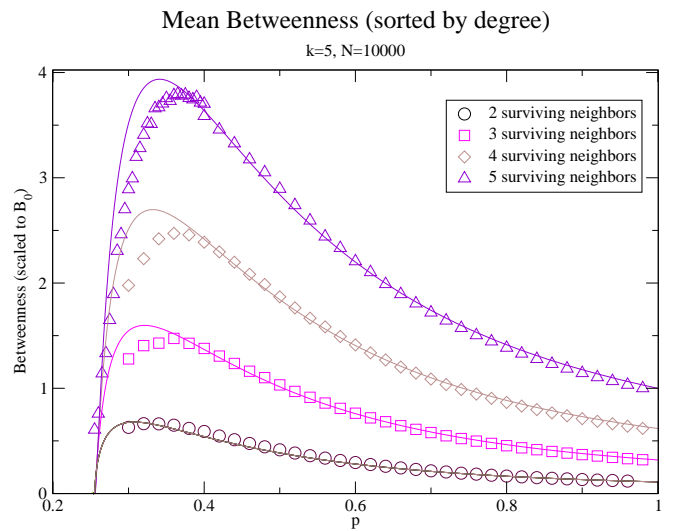}
\caption{Analytical results for the mean betweenness of nodes with different number of neighbors surviving the initial attack as a function of $p$ for a random regular graph. This graph presents the betweenness immediately after the attack without any failure due to overload. The symbols represent simulation results; there are minor discrepancies at very low values of $p$. These discrepancies are discussed later.}
\label{f:Anabtwn}
\end{figure}

In order to better understand the behavior of a graph under a massive attack, we studied the distribution of the betweenness of the nodes for the graphs before the initial attack and just after it (before the cascade of failures takes place).  We start our analysis with the simpler case of RR graphs. Before the initial attack, the betweenness distribution of RR graphs is a sharp Gaussian curve centered around its mean  $\langle B \rangle \approx N\ln (N/k)/\ln (k-1)$\cite{AvgPath}. After the initial attack  the distribution presents a structure in which it is  divided into a number of wider curves, each of which follows a nearly-normal distribution, although with a much larger standard deviation (Fig. \ref{f:betAfter}). The division of the single Gaussian curve into many curves as a result of the initial attack is an important result of this work.  The betweenness of a node $i$ surviving the initial attack is essentially determined by the number of its surviving immediate neighbors, denoted as $\ell_i$.    Specifically, the betweenness is approximately proportional to $\ell_i(\ell_i-1)$, which is similar to, although not identical to, the results found by Goh et al. for scale-free networks \cite{Goh}, in which a scaling relationship between the betweenness of a node and its degree was found.  A theoretical argument for this dependence is given in section \ref{analyticB}, and we present a comparison with our numerical simulations in Fig. \ref{f:Anabtwn}. After the initial attack, our random regular graph loses a fraction $p$ of its nodes, causing the number of surviving first neighbors to vary from node to node.  Most of the  nodes for which all  of the first neighbors survive will have their betweeness increased due to the attack. In contrast, those nodes with neighbors that were destroyed in the initial attack will see their betweenness decrease. 
Accordingly, our results show that the majority of the nodes with all of their neighbors surviving are the nodes whose betweenness will exceed the maximum betweenness and will fail first due to overload, thus driving the phenomena seen in the failure of the network (see Fig. \ref{f:betAfter}).\\

Figure \ref{f:betAfter} provides a lower bound for the value of $p_t(\alpha)$ displayed in
Fig.~\ref{f:pcVsaER} for RR(graphs). Indeed, if we neglect the spread in values of $B(\ell)$, we can assume that $\alpha_0(p)=B(\ell=k,p)/B_0(k)-1 $ gives a good approximation for $\alpha_t(p)$, but due to the spread $\alpha_t(p)>\alpha_0(p)$.
Since both $\alpha_0(p)$ and $\alpha_t(p)$ are decreasing functions of $p$, for sufficiently large $p$, the same is true for the inverse functions. Thus $\alpha_t(p)>\alpha_0(p)$ implies $p_t(\alpha) >p_0(\alpha)$.
More accurate estimates would require the knowledge of the standard deviation of
$B(\ell=k,p)$, which requires additional investigation, which goes beyond the scope of the present paper.

\begin{figure}
\includegraphics[clip,width=\columnwidth]{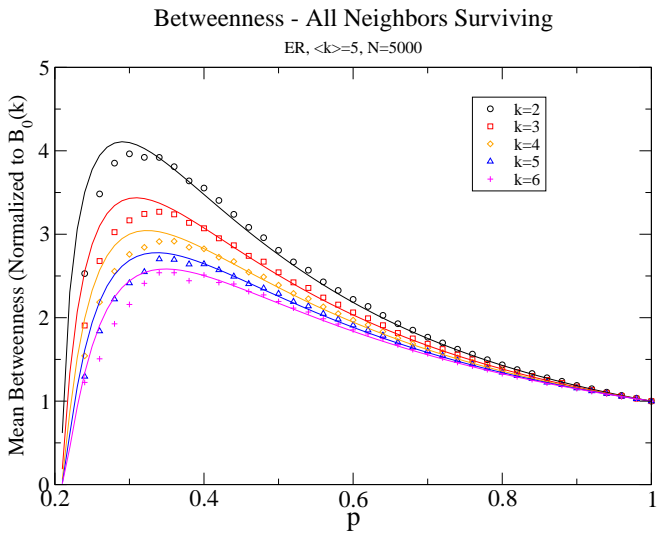}
\caption{Analytical results for the mean betweenness of nodes with all surviving neighbors ($\ell=k$), as a function of $p$ for an Erd\"os-Renyi graph with average degree 5 and size N=5000. All values are normalized to the mean betweenness for that $\ell$ at $p=1$, denoted as $B_0(\ell)$. Note that betweenness rises most for low-degree nodes as $p$ decreases. This graph presents the betweenness without any failure due to overload. The symbols represent simulation results; there are minor discrepancies at very low values of $p$. Note the difference between this graph and Fig. \ref{f:Anabtwn}. Here, we study only nodes in which all their neighbors survive the initial attack, and examine the effect of the original degree on the change in betweenness. In Fig. \ref{f:Anabtwn}, on the other hand, we study nodes with identical original degrees, in which not all neighbors survive.}
\label{f:ERshift}
\end{figure}

In Erd\"os-Renyi (ER) graphs, we find similar catastrophic failures of the network. However, the mechanism of the failure is  slightly different. Because the nodes have different initial degrees, they also have very different initial loads and, thus, different maximum loads. As mentioned (and showed later in section \ref{analyticB}  and  Appendix A),  nodes of lower initial degree will start with lower betweenness and, correspondingly,  lower maximum load.  The initial attack, however, will cause a greater proportional increase in the betweenness of low-degree nodes than in high-degree nodes, as shown in  Fig. \ref{f:ERshift}, provided, in each case, that all neighbors survive.  This will affect the behavior of ER graphs and the way in which they disintegrate. The low-degree nodes fail first (in earlier stages of the cascade and with smaller attacks), causing further fragmenting of the network (see Fig. \ref{f:ERdegfail}). 

If the attack is widespread enough ( $p<p_t$), this fragmentation causes the high-degree nodes to also fail. This multistage phenomenon does not appear to be operative to the same extent in RR graphs;  in those graphs,  relatively few nodes fail before the point where the cascade of failures becomes catastrophic, while in  ER graphs, the decline in the number of nodes in the giant component is more gradual as illustrated in Figure  \ref{f:cascprog}; the low-initial-degree nodes fail first, followed by the hubs. Again, the curve $B(k=2,p)/B(k=2,1)-1$ in
Fig.~\ref{f:ERshift}, provides the lower bound for $p_t(\alpha)$ for the ER graphs.

\section{Features of the Cascades}
\subsection{Progress of Cascade}
\begin{figure}
\includegraphics[clip,width=\columnwidth]{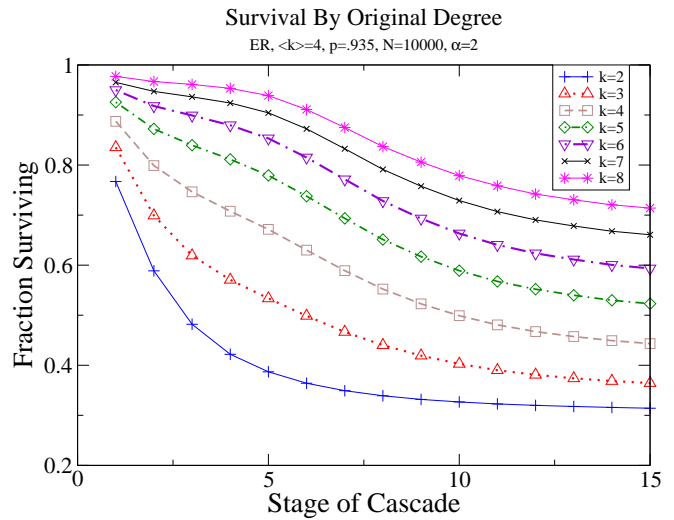}
\caption{Fraction of nodes not failing due to overload at each stage of the cascade, as a function of initial degree. The data presented is for a case of an Erd\"os-Renyi graph where the average degree is 4, $N=10000$, tolerance $\alpha=.2$, and $p_t\approx p=.935$.    Note that nodes with low initial degree fail at a greater proportion, particularly at the early stages of the cascade. The network disintegrates at the end of the cascade; the surviving nodes do not form a giant component.}
\label{f:ERdegfail}
 
\end{figure}

\begin{figure}
\includegraphics[clip,width=\columnwidth]{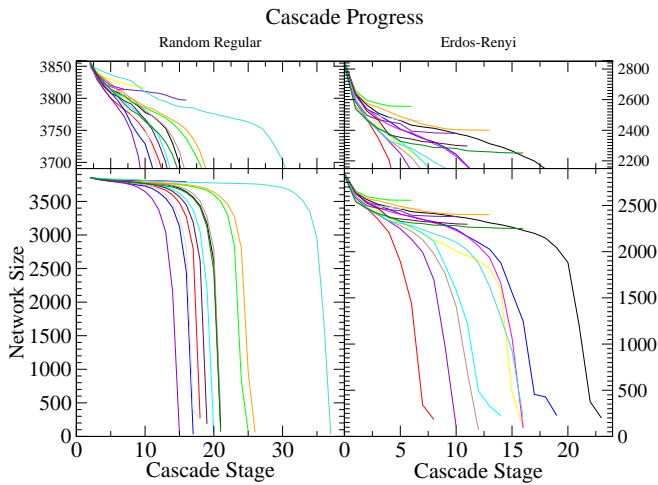}%
\caption{Size of giant component, as the cascade progresses. The left graph corresponds to several realizations of a RR graph with degree 5, $N=5000$, $\alpha=1$, and $p=p_t=.7715$. The right graph corresponds to several realizations of an ER graph with the same parameters and $p=p_t=.603$. Insets above the graph display the first few stages of the cascade in greater detail. Note that for the ER graph, the giant component loses a significant fraction of its size (the low-degree nodes) before catastrophic failure begins, while relatively few nodes fail in the RR graph before the catastrophic portion of the cascade. Nevertheless, both graphs exhibit a first-order transition.}
\label{f:cascprog}
\end{figure}

Immediately after a massive attack near $p_t$, the few nodes with the greatest increase in betweenness fail. As they fail, other nodes increase in betweenness, and also fail. Soon, the network reaches a point of catastrophic failure, in which many nodes fail in each stage of the cascade (Fig. \ref{f:cascprog}). However, there is an important difference between RR and ER graphs. RR graphs have a much more pronounced initial part of the cascade, in which only a few nodes fail.  In ER graphs, instead, we observe faster degradation of the network from the start of the cascades. This is due to the difference in initial degrees; as described, nodes with low initial degrees are most affected by the initial attack. They thus fail first, in the early stages of the cascade. Once they fail, the high-degree nodes fail. This feature is clearly displayed in Fig. \ref{f:ERdegfail}, where the number of nodes surviving each stage of the cascade in an ER graph has been studied as a function of their initial degree.


\subsection{Order of Transition}
\begin{figure}
\includegraphics[clip,width=\columnwidth]{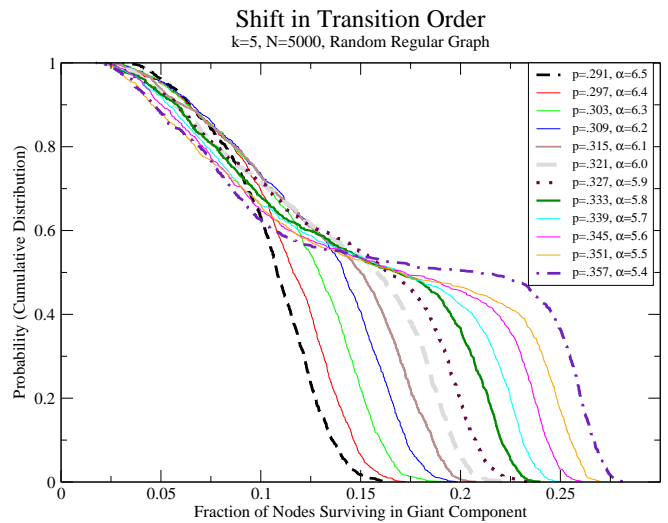}
\caption{Cumulative frequency distribution of the final network size for a variety of $\alpha$'s and their corresponding $p_t$'s.  When $p_t>.32$, where the mean betweenness reaches a maximum (see Fig. \ref{f:Anabtwn}, albeit for a different $N$), the distribution has a plateau, showing a first-order transition in which a midsize network is unstable. When $p_t<.32$, no such plateau exists; when the network begins to disintegrate, it becomes more, not less, stable. The corresponding value of $p_t$ for each $\alpha$ is determined by simulation.}
\label{f:Pc2}
\end{figure}

At high values of $\alpha$, the fragmentation of the network due to the failure of a few nodes can never cause a catastrophic cascade of failures. This is because the betweenness presents a maximum as a function of the fraction of surviving nodes  (see Figs. \ref{f:Anabtwn} and \ref{f:ERshift}). Note that the average betweenness per node in the giant component is $p\hat{N}$L, where $p\hat{N}$ is the number of nodes in the giant component and L is the average path length in the giant component. As fewer nodes survive, the network becomes fragmented, leading to longer path lengths and thus a larger average betweenness. However, at the same time, the fraction of nodes in the giant component decreases, as nodes become isolated due to the widespread destruction. These isolated nodes do not contribute to the betweenness; they do not have paths reaching the nodes in the largest component. Thus, as fewer nodes survive the initial attack, the betweenness decreases. When very few nodes survive, the second effect dominates, and the mean betweenness of nodes decreases with further destruction. In a first-order transition, the original attack causes nodes to fail due to overload, which will cause the mean betweenness to increase. This in turn will cause more destruction; this cascading effect is the scenario that will lead to a first-order transition.\\
However, as long as $\alpha$ is sufficiently high to prevent the network from failing at the point of maximum mean betweenness, the original attack will not cause further failures. Thus, the network will not fail due to a first-order transition. Instead, it will fail due to a second-order transition when the initial attack and associated overload reaches the percolation threshold (which, for random regular graphs, occurs at $\frac{1}{k-1}$.) We define the $p_t$ for this second-order transition as the point where the cascade reaches a maximum in length, analogous to the criterion for first-order transitions.\\
This shift from first-order to second-order scenarios occurs at the value of $\alpha$ where we would expect $p_t$ to equal the fraction of surviving nodes that yields the maximum mean betweenness. That is, if $p_t$ is small enough such that the mean betweenness decreases as $p$ decreases, we will only see a second-order transition, as the fraction of nodes in the giant component decreases to zero due to percolation. In the vicinity of this point, we can see the shift from a first-order transition to a second-order transition as $\alpha$ increases and $p_t$ decreases (See Fig. \ref{f:Pc2}). Thus, the transition between first and second order will occur when $p_t$ is so low that further cascades decrease the average betweenness, and the only failure possible is due to the network reaching  the percolation threshold, and not cascading overload.

\subsection{Size Dependence of the Transition Point}
\begin{figure}
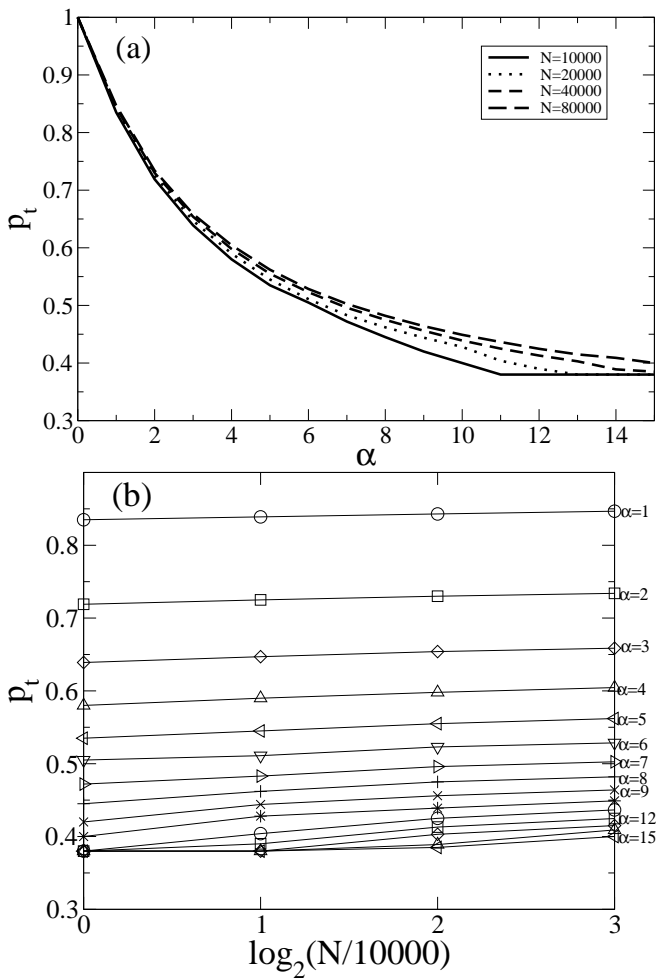

  \includegraphics[clip,width=\columnwidth]{alpha-Pc.eps}
  \includegraphics[clip,width=\columnwidth]{Pc-alpha.eps}
\caption{(a) Dependence of the transition point $p_t(\alpha)$ on the tolerance $\alpha$ for different system sizes, $N=10000,20000,40000,80000$. For large tolerance the curves for all system sizes converge from above to the critical point $p_c\approx0.38$, at which the bimodal first order transition change to a unimodal second order transition. (b) Dependence of the transition point $p_t(N)$ on the system size for different tolerances $\alpha$. One can see almost linear increase of $p_t$ with $\ln N$. } 
\label{f:size}

\end{figure}

The logarithmic dependence of the betweenness on the system size produces a strong logarithmic dependence of the transition point $p_t$ on the size of the system $N$, and
also changes the location of the critical point $p_c$ at which the first order phase transition switches to a second order percolation transition.
Fig. \ref{f:size} (a) shows the behavior of $p_t(\alpha)$ for RR graphs ($k=4$), for  $N=10000,20000,40000,80000$. One can see that the larger networks becomes
more vulnerable than the smaller ones. This phenomenon is similar to the one observed in \cite{Lowinger} for high dimensional interdependent lattices. For all values of $N$, the curves $p_t(\alpha)$ approach the critical point $p_c\approx 0.38$ from above. Note that this value is almost independent of $N$. Since for larger $N$, these curves are going higher, they reach the critical point
 at larger values of $\alpha$. Figure \ref{f:size} (b) shows the increase of $p_t$ as
a function of the  size of the  system for different $\alpha$. One can see that $p_t(N)$
grows approximately linearly with $\ln(N)$ above the critical point value $p_c$.

\section{Analytical Calculations of the betweenness}
\label{analyticB}
In order to estimate the betweenness, we need to define an exterior and a shell. We define $x_n$ as the fraction of nodes more than $n$ nodes away from a central node and $y_n$ as the fraction of nodes that are exactly $n$ nodes away. We further define $x_\infty$ as the fraction of nodes that are isolated from the giant component of the network. These are all expressed as fractions of $\tilde{N}\equiv pN$, the size of the decimated network.
By definition,
\begin{equation}
y_{n}=x_{n-1}-x_n.
\label{y_n}
\end{equation}
In the rest of this section, we will illustrate our findings for the case of  random regular graphs, and will collect results from Erd\"os-Renyi graphs in Appendix A.\\
Following  \cite{Buldyrev24}, we use the relationship 
\begin{equation}
x_{n+1}=G_0(G_1[G^{-1}_0(x_n)]),
\end{equation}
where $G_0$ is the generating function of the network, $G_1\equiv \frac{G_0'(x)}{G_0'(1)}$ is the generating function of the branching proces, and $G_0^{-1}$ is the inverse of $G_0$.\\
In the case of RR graphs this expression becomes,
\begin{equation}
x_{n+1}=(x_n^{\frac{k-1}{k}})^k=x_n^{k-1}
\label{x_n+1}
\end{equation}

It is known \cite{Newman} that if a random fraction $(1-p)$ of the nodes are destroyed in a network which initially had a generating function given by $G_0(x)$, the generating function of the decimated network becomes $G_0(1-p+px)$ for the same function $G_0$. Thus, for a decimated random regular network with initial degree $k$, we obtain
\begin{equation}
x_{n+1}=(1-p+px_n^{\frac{k-1}{k}})^k.
\label{e:x0}
\end{equation}
This relationship allows us to create shells of nodes around a central node which ends up with $\ell$ surviving neighbors after the initial attack. Setting $y_1(\ell)=\ell / \tilde{N}$ by definition,  and doing  a Taylor expansion of $x_0(\ell)$ around 1, and using Eqs. (\ref{y_n}) and (\ref{e:x0}) for the case $n=0$,  we obtain for the case of RR graphs 
\begin{equation}
x_0(\ell)=1-\frac{\ell}{\tilde{N}[p(k-1)-1]},
\label{e:initcond}
\end{equation}

\begin{figure}
  \includegraphics[clip,width=.5\columnwidth]{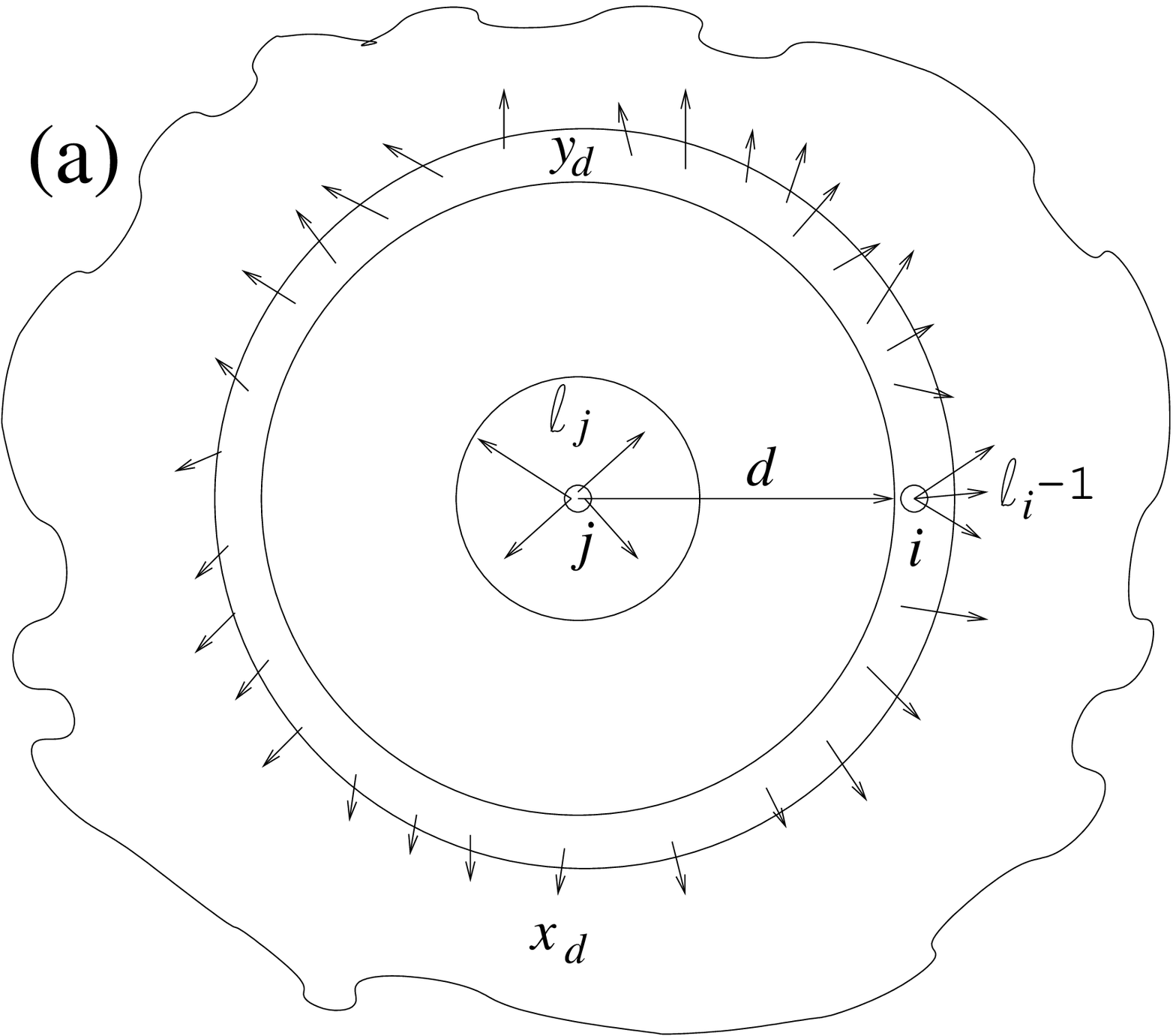}%
  \includegraphics[clip,width=.35\columnwidth]{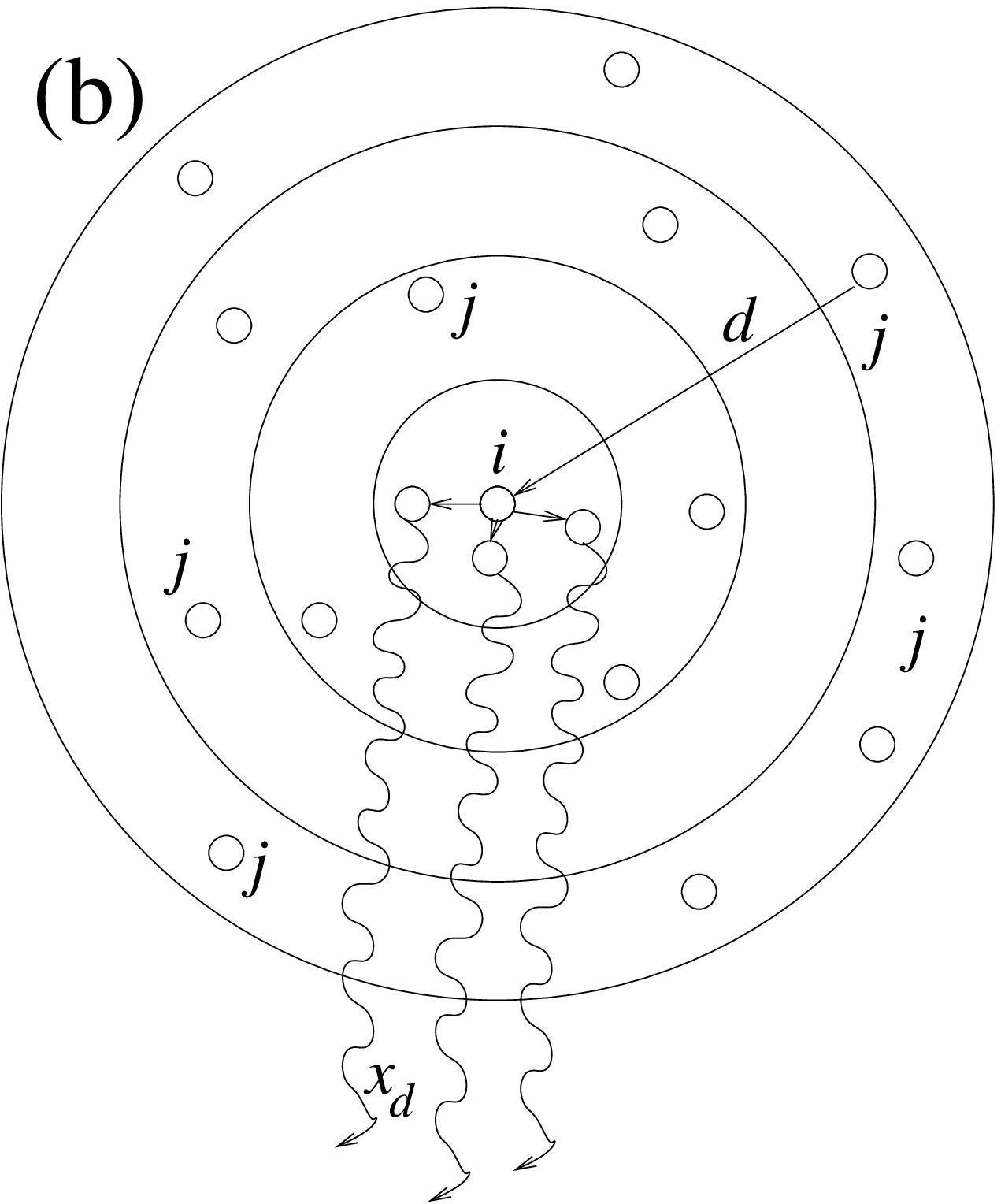}%
  \caption{A schematic illustration of the calculation of the average betweenness.
    (a) derivation of Eq. (\ref{e:BDtilda}).(b) derivation of Eq. (\ref{e:B}).
    In panel (a), the original node $i$ with degree $\ell_i$ is placed in the $d$-th shell of node $j$, which is depicted at the center. The exterior of the $d$-th shell $x_d$ of node $j$ with degree $\ell_j$, $x_d$, is connected to the $d$-th shell $y_d$ by links depicted by arrows emanating from the $d$-th shell;  $\ell_i-1$ of those links belong to node $i$. We assume that each of these links from shell $y_d$ carries on average equal amount of shortest paths, from $x_d$ to node $j$. In panel (b), the original node $i$ for which the total betweenness is calculated is placed at the center, and the contribution of node $j$ to its betweennes must be summed for all nodes $j$ and shells $d$. The connections of node $i$ to the exterior of the $d$-th shell of node $j$, $x_d$ are shown by wavy lines.       
  } 
\label{f:schematic-a}
\end{figure}

With these equations, we are now in a position  to calculate the betweenness of a node $i$ with $\ell_i$ surviving neighbors. In order to proceed, we first study the contribution to the betweenness from paths that leave another node $j$ and travel through $i$, where $j$ is a distance $d$ away from $i$ (Fig.~\ref{f:schematic-a}) To do so, we recreate the graph, using $j$ as a new "central node", around which we build shells. When we recreate the graph around $j$, our original node $i$ is in a shell a distance $d$ away, and $\tilde{N}[x_d(\ell_j)-x_\infty]$ of the nodes in the network belong to the giant component, but are farther away from $j$ than the original node $i$ is. The shortest path between $j$ and any of these nodes in the $d$-exterior (of $j$) must pass through (or originate in) the $d+1$-shell, and then travel from there through a link to the $d$-shell. We will assume that each of these links between the $d$ and $d+1$ shells (depicted as arrows in Fig.~\ref{f:schematic-a}(a)) carries an equal amount of traffic.\\
$(\ell_i-1)$ of these links branch out of the original  node $i$; while an average of $p(k-1)[\tilde{N}y_d(\ell_j)-1]$ links branch out from the other nodes (different from $i$) in the $d$-shell. Thus, the contribution to the betweenness of $i$ due to a \underline{single} node $j$  a distance $d$ away is
\begin{equation}
\tilde{B_d}(\ell_i,\ell_j)=\tilde{N}[x_d(\ell _j)-x_\infty]\frac{(\ell _i-1)}{p(k-1)[\tilde{N}y_d(\ell_j)-1]+\ell_i-1},
\label{e:BDtilda}
\end{equation}
In this expression, the numerator of the fraction is the number of links that branch out from the $i$ node and the denominator is the total number of links that branch out from all the nodes in the $d$-shell of $j$. \\
This expression contains a slight error; the actual number of links that branch out of  $j$'s $d$-shell is a random variable with a mean at the value given. For computational simplicity, we treat the mean of the fraction as the fraction of the means. This simplification causes errors at low values of $p$, where there is a greater variation in the denominator (Fig. \ref{f:Anabtwn})). Taking a second order Taylor expansion of Eq. (\ref{e:BDtilda}), we find a correction factor of
\begin{equation}
\tilde{B_d}(\ell_i,\ell_j)\sigma_{y_d(\ell_j)}^2 / y_d(\ell_j)^2
\end{equation}
While we have not calculated $\sigma_{y_d(\ell_j)}^2$ directly, the approximation should be noted.\\

With our value of $\tilde{B_d}(\ell_i, \ell_j)$ for the betweenness of a node $i$ due to paths leaving a single node $j$, we now make an identical argument for each node $j$ a distance $d$ away from $i$, for each value of $d$ (Fig.~\ref{f:schematic-a}). In order to do that, we must perform a sum over all $j$. This requires calculating the probability distribution of $\ell_j$ for a given node $j$. Node $j$ will have $\ell_j$ surviving neighbors with probability \cite{Buldyrev24}
\begin{equation}
\tilde{P}(\ell_j,d)=P(\ell_j)[G_0^{-1}(x_{d-1})^{\ell_j}-G_0^{-1}(x_{d})^{\ell_j}]/y_d,
\end{equation}
where $P(\ell_j)$ is the overall fraction of nodes in the network with $\ell_j$ surviving neighbors, or $C_k^{\ell_j} (1-p)^{k-\ell_j} p^{\ell_j}$. Summing over all $d$ and all $j$ we find that the total betweenness is

\begin{eqnarray}
B(\ell_i)&=\sum_{d=1}^\infty[ \tilde{N}y_d(\ell_i)\sum_{\ell_j=1}^k\tilde{B_d}(\ell_i,\ell_j)\tilde{P}(\ell_j,d)]\\
&\equiv \sum_{d=1}^\infty[\tilde{N}y_d(\ell_i)\langle \tilde{B_d}(\ell_i)\rangle]
\label{e:B}
\end{eqnarray}
This is the closest approximation we have for the betweenness of a node and the equation we use in Fig. \ref{f:Anabtwn}. Note that $\tilde{B}$ is proportional to $\ell-1$ and $y_d$ is approximately proportional to $\ell$, giving us the  $\ell(\ell-1)$ dependence of the betweenness discussed in section \ref{gen_res}\\
For large $\tilde{N}y_d$ or $p(k-1)\approx \ell_i-1$, we can simplify the denominator in Eq. (\ref{e:BDtilda}) and average over all $\ell_j$ (again, introducing a slight error term due to equating the fraction of the averages with the average of the fractions),  leading to an average value of $\tilde{B_d}$ for each $l_j$:
\begin{equation}
\langle \tilde{B_d}(\ell_i)\rangle\approx (\langle x_d\rangle -x_\infty)\frac{(\ell_i-1)}{p(k-1)\langle y_d\rangle }
\label{e:avgB}
\end{equation}
and thus, combining Eqs. (\ref{e:B}) and (\ref{e:avgB}),
\begin{equation}
B(\ell_i)\approx \tilde{N}\sum_{d=1}^\infty\frac{y_d(\ell_i)}{\langle y_d\rangle}\frac{(\ell_i-1)}{p(k-1)}(\langle x_d\rangle -x_\infty)
\label{e:B2}    
\end{equation}

Note that for $\ell_i\approx k$, the two fractions are near unity, and we are left with the intuitive result that the mean betweenness will be the average path length, which we will call $L$, multiplied by the network size. This follows from the observation that  $\sum \langle x_i\rangle$ can be written telescopically as $\sum (\langle x_i\rangle - \langle x_{i+1}\rangle )\times i \equiv \sum i \times \langle y_i\rangle$.  For other values of $\ell_i$, note that for small $d$, $\frac{y_d(\ell _i)}{\langle y_d\rangle}\approx\frac{\ell_i}{pk}$, and thus, for $\ell_i=k_{max}$ (that is, the betweenness of a node in an RR graph with all of its neighbors surviving),
\begin{equation}
B(\ell_i)\approx \frac{p\hat{N}\ell_i (\ell_i-1)L}{pk  p(k-1)}= \frac{p\hat{N}L}{p^2}=\frac{\hat{N}L}{p}
\end{equation}
where $p\hat{N}$, as earlier, is $pN(1-x_{\infty})$, or the total number of nodes in the giant component. We have explicitly written out this prefactor here, to more clearly identify the dependence of our result on $p$
Using results from \cite{AvgPath}, we finally obtain for the betweenness of a node in a random regular graph of degree $k$ 
\begin{equation}
B(k)\approx \frac{\hat{N}}{p} (\frac{\ln(p\hat{N}/pk)}{\ln[p(k-1)]})= \frac{\hat{N}}{p} (\frac{\ln(\hat{N}/k)}{\ln[p(k-1)]})
\label{finBRR}
\end{equation}

While this analysis has been illustrated with random regular graphs, the results also hold true for any random graph, \textit{mutatis mutandis}, and in Appendix A we reobtain them for Erd\"os-Renyi graphs. \\

These results are confirmed within 5\%, for $p\approx1$, where this approximation is most accurate. Note that the betweenness of the nodes with all of their original neighbors intact increases as approximately $1/p$, which is the primary cause of network failure. \\

\section{Localized Attack}
\begin{figure}
\includegraphics[clip,width=\columnwidth]{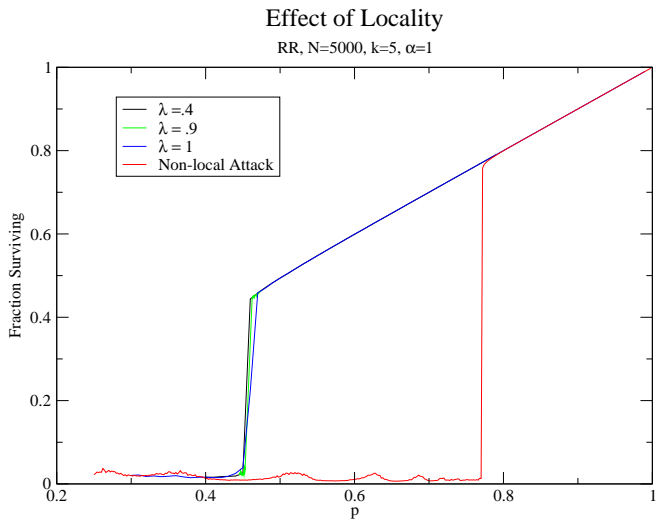}%
\caption{Size of the final giant component as a function of $p$. This graph shows the effect of a local attack, and the effect of several values of $\lambda$, a measure of the locality of the attack. For the case of $\lambda=1$, a random node and the nearest $1-p$ nodes are destroyed. Although this can lead to a cascade of failures, such a cascade happens only at a relatively low value of $p$. Although the exact value of $\lambda$ has little effect on the vulnerability, all local attacks are much less effective than non-local attacks. This graph shows the results from a random regular graph of $N=5000, k=5$, and $\alpha=1$.
}
\label{f:local1}
\end{figure}

\begin{figure}
\includegraphics[clip,width=\columnwidth]{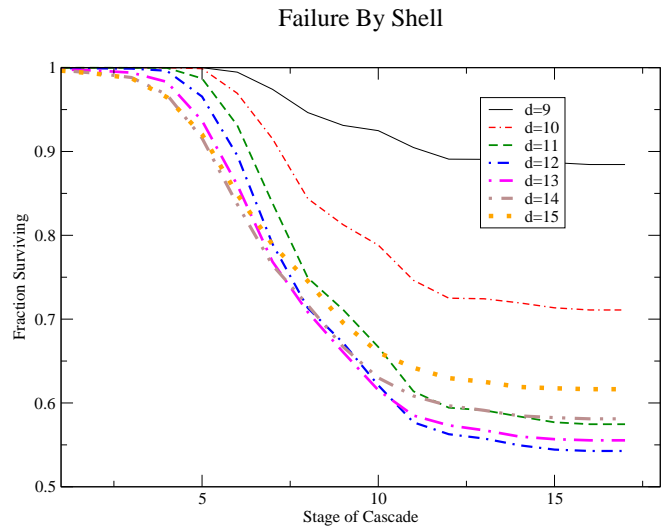}%
\caption{Fraction of nodes not overloaded, as the cascade progresses, for different values of $d$, the distance from the central node. Note that the nodes closest to the center of the destruction are the last to be destroyed; because their neighbors are destroyed in the original attack, their betweenness decreases. Conversely, the nodes farthest from the center of the destruction fail first; not only do all of their immediate neighbors survive, more of their distant neighbors survive, increasing their betweenness beyond the values that we have calculated. The results included are from a random regular graph, $N=10000$, $k=3$, $\alpha=.2$, $p=p_t=.908$, and $\lambda=1$. The first 8 shells around the central node are completely destroyed in the initial attack and thus are not shown on the graph.
}
\label{f:local2}
\end{figure}

According to our theory, the network becomes vulnerable because of the variation in the number of surviving neighbors in the nodes. Thus, a localized attack, in which most nodes' neighbors are unaffected, will be less effective than a comparable random attack. We implement this local attack by selecting a random node in the network. This node is destroyed, and the destruction spreads along the network's links to each of its neighbors with probability $\lambda$, a measure of the locality of the attack. When the attack reaches a neighboring node, that node is also destroyed, and the damage spreads to its neighbors with the same probability $\lambda$. Once $1-p$ nodes are destroyed, the initial attack ceases, and we begin to evaluate the failure of nodes due to overload. This allows us to interpolate between the case of a totally random attack $ \lambda= 1-p$  and a totally localized attack with $ \lambda= 1$. Simulations show (Fig \ref{f:local1}), as we expected, that a network can survive a local attack, even when a random attack of the same strength would have caused the network's collapse.
Other models have assumed that the destruction wrought by an initial attack often spreads to the attacked nodes' neighbors first (\cite{Brummitta,Wang3}). However, our results show the opposite in this model; the nodes closest to the destruction are the least likely to be overloaded. This is confirmed by the progression of the cascade in the case of a localized attack. After the initial attack, the nodes farthest from the center of the destruction are the first nodes to fail. Only after they fail do the inner nodes, near the nodes destroyed in the initial attack, fail (Fig. \ref{f:local2}). This difference emphasizes the difference between the random networks and the networks embedded in space for which the opposite effect is observed.\cite{Havlin}.
\section{Conclusion}
We have studied, both computationally and analytically, the effects of
widespread attacks on networks that are susceptible to fail due to
betweenness overload.  We study the fraction of survived nodes at the
end of the cascade $p_f$ as function of the strength of the initial
attack, measured by the fraction of nodes $p$, which survive the
initial attack for different values of tolerance $\alpha$ in random
regular and Erd\"os-Renyi graphs. We find the existence of first order
phase transition line $p_t(\alpha)$ on a $p-\alpha$ plane, such that
if $p <p_t$ the cascade of failures lead to a very small fraction of
survived nodes $p_f$ and the giant component of the network
disappears, while for $p>p_t$, $p_f$ is large and the giant component
of the network is still present. This feature of the cascading
failures is similar to the phenomenology found in other models of
cascading failures: i.e. bootstrap
percolation\cite{Watts,Baxter2010,Baxter2011,Gleeson} k-core
percolation\cite{Baxter2011} and mutual percolation in interdependent
networks\cite{Buldyrev,DiMuro2016,DiMuro2017}.  Exactly at $p_t$ the
function $p_f(p)$ undergoes a first order discontinuity. We find that
the line $p_t(\alpha)$ ends at critical point $(p_c,\alpha_c)$, in which
the cascading failures are replaced by a second order percolation
transition. We analytically find the average betweenness of nodes with
different degrees before and after the initial attack, investigate
their roles in the cascading failures, and find a lower bound for
$p_t(\alpha)$. The dynamics of cascading failures indicates the
existence of a latent period of cascading failures, during which only
a few overloads occur at each stage of the cascade. This latent period
is more pronounced in RR graphs than in ER graphs. A similar latent
period is present in a more realistic model of overloads in the power
grid based on a direct current approximation (DC)\cite{Spiewak}. Another similarity between the Motter and Lai model and the DC model of the power grid is a complete clusterization of the network at the end of the cascade. In both models, the giant cluster remains to be the most vulnerable until the last stages of the cascade. In the small clusters of the Motter and Lai model, nodes have low betweenness and, thus, do not suffer from overloads, without adding to global transport in the network. In the power grid model, small self-sustaining islands are likely to survive, because local transmission lines connecting neighboring consumers and producers are less likely to develop overloads than lines connecting distant
parts of the network, which may develop a huge imbalance of production and consumption.     

Our main finding is that the degree of a node is the primary determinant of its betweenness, and thus its risk of overloading. This shows the fragility of nodes with many surviving neighbors, and of nodes with low initial degrees in non-regular networks. This knowledge can be used to stop cascades in their track, or to easily identify the most vulnerable nodes. This result has led to new insights on the critical point, at which the transition shifts from first-order to second-order, and the effect of the degree of the network, the degree distribution, the size of the network, and the tolerance on the stability of the network. 
We also study the difference of cascading failures caused by local attacks and random attacks on randomly connected networks. We find that localized attacks are less destructive than random attacks, which is opposite to the behavior of spatially embedded networks \cite{Havlin}.\\
\section{Acknowledgements}
Our research was supported by HDTRA1-14-1-0017. SVB acknowledge the partial support of this research through the Dr. Bernard W. Gamson Computational Science Center at Yeshiva College.

\section{Appendix A- Calculation of Betweenness for the Erd\"os-Renyi model }

For the case of Erd\"os-Renyi graphs, the expression equivalent to (\ref{e:x0}) becomes 
\begin{equation}
x_{n+1}=e^{p\langle k\rangle(x_n-1)}.
\end{equation}
and the Taylor expansion equivalent to (\ref{e:initcond}) is
\begin{equation}
x_0(\ell)=1-\frac{\ell}{\tilde{N}[p\langle k\rangle-1]}.
\end{equation}
The shell analysis for the contribution of a single node to the betweenness  (equivalent to Eq. (\ref{e:BDtilda}) yields
\begin{equation}
\tilde{B_d}(\ell_i,\ell_j)=\tilde{N}[x_d(\ell _j)-x_\infty]\frac{(\ell _i-1)}{p\langle k\rangle [\tilde{N}y_d(\ell_j)-1]+\ell_i-1},
\end{equation}

while equation (\ref{e:B}) for the total betweenness  still remains valid. 

In this case the approximations analogous to Eqs. (\ref{e:avgB})  and (\ref{e:B2}) are
\begin{equation}
\langle \tilde{B_d}(\ell_i)\rangle\approx (\langle x_d\rangle -x_\infty)\frac{(\ell_i-1)}{p(k)\langle y_d\rangle }
\label{e:aERvgB}
\end{equation}

\begin{equation}
B(\ell_i)\approx \tilde{N}\sum_{d=1}^\infty\frac{y_d(\ell_i)}{\langle y_d\rangle}\frac{(\ell_i-1)}{(p\langle k \rangle)}(\langle x_d\rangle -x_\infty)
\label{e:ERB2}
\end{equation}

And finally, the result obtained using \cite{AvgPath}, equivalent to Eq. (\ref{finBRR}) becomes

\begin{equation}
B(k)\approx \frac{\hat{N}}{p}(\frac{\ln(\hat{N}/\langle k\rangle)}{\ln(p \langle k \rangle)})
\end{equation}
for Erd\"os-Renyi graphs, where $\langle k \rangle $ represents the average degree before the attack.

\end{document}